\documentclass[12pt]{iopart}
\pdfoutput=1
\usepackage{iopams}
\usepackage{graphicx}
\usepackage{hyperref}
\usepackage{graphicx}
\usepackage{color}
\usepackage[amssymb]{SIunits}

\begin{document}

\title{Electron interferometry in integer quantum Hall edge channels}

\author{J Rech, C Wahl, T Jonckheere and T Martin}
\address{Aix Marseille Universit\'e, Universit\'e de Toulon, CNRS, CPT, UMR 7332, 13288 Marseille, France}
\ead{jerome.rech@cpt.univ-mrs.fr}

\begin{abstract}

We consider the electronic analog of the Hong-Ou-Mandel interferometer from quantum optics. In this realistic condensed matter device, single electrons are injected and travel along opposite chiral edge states of the integer quantum Hall effect, colliding at a quantum point contact (QPC). We monitor the fate of the colliding excitations by calculating zero-frequency current correlations at the output of the QPC.
In the simpler case of filling factor $\nu=1$, we recover the standard result of a dip in the current noise as a function of the time delay between electron injections. For simultaneous injection, the current correlations exactly vanish, as dictated by the Pauli principle.
This picture is however dramatically modified when interactions are present, as we show in the case of a filling factor $\nu=2$. There, each edge state is made out of two co-propagating channels, leading to charge fractionalization, and ultimately to decoherence. The latter phenomenon reduces the degree of indistinguishability between the two electron wavepackets, yielding a reduced contrast in the HOM signal.
This naturally brings about the question of stronger interaction, offering a natural extension of the present work to the case of fractional quantum Hall effect where many open and fascinating questions remain.


\end{abstract}



\maketitle


\section{Introduction}

Electron quantum optics (EQO) aims at transposing quantum optics experiments, allowing for the controlled preparation, manipulation and measurement of single electronic excitations in ballistic quantum conductors. 
One may expect fundamental departures from their photon counterpart, as electrons are not only subject to Coulomb interactions, but they also obey the fermionic statistics. While the control of single photons - a key ingredient of quantum optics experiments - was mastered long ago~\cite{clauser74}, such a feat was only achieved recently in condensed matter devices~\cite{feve07,leicht11,mcneil11,hermelin11,giblin12,dubois13}.

High-mobility 2D electron gases are a perfect testbed for conducting EQO experiments as several building blocks of quantum optics can readily be recreated in this context. First, the phase-coherent ballistic propagation of electrons is ensured by chiral edge states of the integer quantum Hall effect (IQHE). After propagation, these electrons collide at a quantum point contact (QPC), a tunable tunnel barrier mimicking a beamsplitter. The only missing ingredient finally appeared recently in the form of an on-demand single electron source (SES), opening the way to all sorts of interference experiments~\cite{feve07}.

\begin{figure}
\begin{center}
a)
\includegraphics[height=3.5cm]{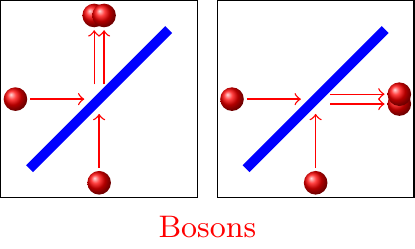}
\hspace{1.5cm}
b)
\includegraphics[height=3.5cm]{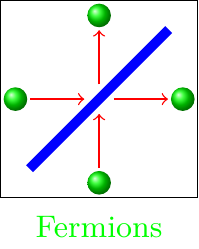}
\end{center}
\caption{
Schematic view of the HOM collision where two incoming objects scatter on a beam splitter (in blue). In the bosonic case (left), two bosons exit in the same output leading to two different possible outcomes and the vanishing of the coincidence count between the two output channels. In the fermionic case (right), the two fermions exit in different output channel, leading to a unique possible outcome and thus to a vanishing of the current fluctuations at the exit from the beam splitter. This shows the clear link between the outcome of the HOM interferometry experiment and the bunching or antibunching properties of the colliding objects, and thus their statistics.
}
\label{fig:bunching}
\end{figure}

Among those, the Hong-Ou-Mandel~\cite{hong87, beugnon06} (HOM) interferometer is a celebrated tool of quantum optics. It allows to probe the degree of indistinguishability of two photons sent on a beamsplitter, by measuring the coincidence rate between the two output channels.
When identical photons are sent on the two input channels of a beamsplitter, and collide at the same time, they exit in the same outgoing channel, showing a sudden vanishing of the output coincidence rate (see Fig.~\ref{fig:bunching}). This bunching phenomenon is a direct consequence of the bosonic statistics. Moreover, measuring this dip gives access to the size of the photon wavepacket and the time delay between photon emissions.

The electronic analog of the HOM experiment in condensed matter goes beyond the simple transposition of an optics setup as several major differences exist between photons and electrons. In particular, electrons differ because of the presence of the Fermi sea and the possibility of creating electron vacancies - i.e. holes - but also electrons are susceptible to interact with each other leading to new and interesting effects that have no equivalent with photons. This device has so far eluded a complete theoretical description~\cite{burkard03,giovannetti06,feve08,olkhovskaya08,moskalets11}.

Here we study, from a theoretical standpoint, the outcome of this electronic HOM interferometry experiment, where two independently emitted electrons travel along counter-propagating opposite edge states and meet at a QPC, in the integer quantum Hall regime at both filling factor $\nu =1$ and $\nu=2$. The latter case allows us to not only investigate the effect of Coulomb interactions along the propagation but also to provide a theoretical framework for recent experimental results obtained at $\nu>1$~\cite{bocquillon13}.

This article is organized as follows. In Sec.~\ref{sec:nu1}, we present the derivation and main results in the $\nu=1$ IQH case, insisting on the two possible cases of two-fermion interferences. We then develop in Sec.~\ref{sec:nu2} the formalism allowing us to incorporate inter-channel interaction, and argue that an interaction-based decoherence scenario can explain the recent experimental results. In Sec.~\ref{sec:upon}, we present the challenges to overcome in order to extend the idea of an HOM interferometry setup to the more complicated but fascinating case of the fractional quantum Hall regime. Finally Sec.~\ref{sec:conclusion} is devoted to the conclusion.

\section{HOM interferometer at filling factor $\nu=1$} \label{sec:nu1}

\subsection{Setup}

\begin{figure}
\begin{center}
\includegraphics[height=4.5cm]{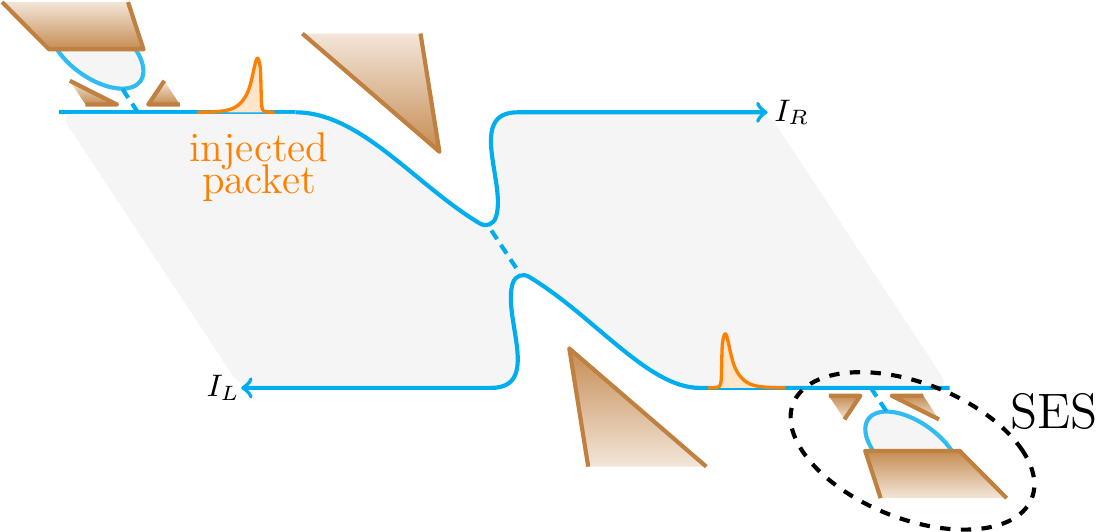}
\end{center}
\caption{
Schematic view of the HOM setup. Two counter-propagating edge states, each equipped with a single electron source, are coupled at a quantum point contact. Correlations between the output currents $I_R(t)$ and $I_L(t)$ are calculated as a function of the time difference between electron emissions.
}
\label{fig:setup}
\end{figure}

The electronic HOM interferometer involves two counter-propagating chiral edge states which meet at a QPC. In the case of filling factor $\nu=1$, each edge is made of a single channel for electrons to propagate. Single electrons can be emitted into the system with a tunable time difference thanks to single electron sources which are connected to each incoming edge states. The current is measured at each output channel, and one can compute current correlations which are conveniently expressed as a function of the time delay between injected electrons. A schematic setup is presented in Fig.~\ref{fig:setup}.

Valuable physics is encoded in the noise properties of the system, and in particular the quantity of interest for us is the zero-frequency current correlations at the output of the QPC, which read
\begin{equation}
S_{RL}^{\rm out} = \int dt dt' \left[ \langle I_{R}^{\rm out} (t) I_{L}^{\rm out} (t') \rangle - \langle I_{R}^{\rm out} (t) \rangle \langle I_{L}^{\rm out} (t') \rangle \right],
\end{equation}
where $I_{R}^{\rm out} (t)$ and $I_{L}^{\rm out} (t)$ are the currents in the two output channels ($R/L$ being right- and left-movers).

\subsection{Formalism} \label{sec:nu1formalism}

Using the linear dispersion of the chiral edge states, it turns out that the currents only depend on $x - v_F t$ ($v_F$ being the Fermi velocity along the edge). It follows that the zero-frequency current correlations can be expressed in terms of the current at the immediate output of the QPC.  The latter is written in terms of the outgoing fermionic fields as
\begin{equation}
I_\mu^{\rm out} (t) = - e v_F  : \left(\psi_\mu^{\rm out} (t)\right)^\dagger  \psi_\mu^{\rm out} (t) : ,
\end{equation}
where $:...:$ stands for normal ordering.

The quantum point contact is modeled using a scattering matrix~\cite{grenier11} which couples the outgoing fermionic fields to the incoming ones according to
\begin{equation}
\left( 
\begin{array}{c}
\psi_{L}^{\rm out} (t) \\ 
\psi_{R}^{\rm out} (t) 
\end{array}
\right)
 = 
\left( 
\begin{array}{cc}
\sqrt{\mathcal{T}} & -i \sqrt{\mathcal{R}} \\ 
-i \sqrt{\mathcal{R}} &  \sqrt{\mathcal{T}}
\end{array}
\right)
\left( 
\begin{array}{c}
\psi_{L}^{\rm in} (t) \\ 
\psi_{R}^{\rm in} (t) 
\end{array}
\right) ,
\end{equation}
where the transmission and reflexion probabilities are given by $\mathcal{T}$ and $\mathcal{R} = 1 - \mathcal{T}$ respectively.

This allows us to write the outgoing currents in terms of incoming fermion fields, whose correlation functions are known
\begin{eqnarray}
I_{R}^{\rm out}(t) &=  {\mathcal T} I_R(t) +  \mathcal{R} I_L(t) - i e 
           \sqrt{\mathcal{R} \mathcal{T}}(\psi^{\dagger}_R \psi_L - \psi^{\dagger}_L \psi_R)(t)   \\ 
I_{L}^{\rm out}(t) &=  {\mathcal R} I_R(t) + {\mathcal T} I_L(t) + i e 
           \sqrt{\mathcal{R} \mathcal{T}} (\psi^{\dagger}_R \psi_L - \psi^{\dagger}_L \psi_R)(t)      
\end{eqnarray}
where for notational convenience, we dropped the "in" superscript. Substituting these expressions back into the definition of the current noise, one is left with
\begin{eqnarray}
S_{RL}^{\rm out} = \mathcal{R} \mathcal{T} \int dt dt' & \left\{ \langle I_R(t) I_R(t') \rangle + \langle I_L(t) I_L(t') \rangle \right. \nonumber \\
&  \left. - e^2 \left[ \langle \psi_R(t) \psi_R^\dagger(t') \rangle \langle \psi_L^\dagger (t) \psi_L (t') \rangle + {\rm H.c.} \right]  \right\} .
\label{eq:noisein} 
\end{eqnarray}
In order to perform analytic calculations, we need to resort to a simplified model for the emission of electrons. We thus consider the injection of single electrons with a given exponential wavepacket added to each edge, in close similarity to the state of the system when the single electron source is operated in its optimal regime.

The states describing each edge are then given by the application of a fermionic operator with a given envelope, namely
\begin{equation}
| \Psi^{e/h}_{\mu} \rangle = \int dx \; \phi^{e/h}_{\mu}(x) \; \psi^{e/h}_{\mu}(x) \; | 0 \rangle ,
\label{eq:injwp}
\end{equation}
where $| 0 \rangle$ stands for the Fermi sea at temperature $\Theta$, and $\psi^e = \psi^\dagger$ corresponds to injecting a single electron while $\psi^h = \psi$ corresponds to a single hole. All averages in Eq.~(\ref{eq:noisein}) have to be evaluated over this prepared state, which corresponds to the state of the system after injection on a given edge.

\subsection{Symmetric electron-electron collisions}

We first consider the case of a symmetric electron-electron collision, where identical electronic wavepackets (corresponding to $\phi_R (x) = \phi_L (x) = \phi (x)$) reach the QPC with a time difference $\delta t$. Working out the algebra, the expression for the current cross-correlations reduces to
\begin{equation}
\frac{S_{RL}^{\rm out}(\delta t)}{2 \mathcal{S}_{HBT}} =
1 - \left| \frac{\int_0^\infty dk
  |\phi(k)|^2 e^{-i k \delta t} (1-f_k)^2}
  {\int_0^\infty dk  |\phi(k)|^2 (1-f_k)^2} \right|^2 ,
\label{eq:symHOM}
\end{equation}
where $\phi (k)$ is the wavefunction in momentum space related to the injected electron, and $f_k = 1/(1+e^{(k-k_F)/\Theta})$ corresponds to the Fermi distribution. Here, we decided to normalize the results of the HOM interferometry by the noise associated with a single electron scattering at the QPC, the so-called Hanbury-Brown Twiss (HBT) contribution $\mathcal{S}_{HBT}$~\cite{hanburybrown57,hanburybrown58}, which we have to count twice to account for the two injected electrons.

Already at this stage, the expression for the noise shows some interesting behavior in a few limiting cases, independently of the actual wavepacket emitted. First, for large values of the time difference $\delta t$ between electrons, the ratio $\frac{S_{RL}^{out}(\delta t)}{2 \mathcal{S}_{HBT}}$ saturates to 1. Indeed, if $\delta t$ is larger than the spread of the wavepacket in time, the two electrons no longer interfere at the QPC and the noise reduces to the contributions from the two electrons taken independently. On the opposite, for simultaneous injections, the noise shows a dip and vanishes exactly at $\delta t = 0$, as expected from Pauli principle since there is only one possible outcome for the two outgoing electrons. 

More specifically, we now focus on the case of exponential wavepackets. These correspond to the behavior of the experimental single electron source in its optimal regime of operation. Indeed in this case, the source can be viewed as a single energy level coupled to a continuum and driven by a square voltage centered around the Fermi energy. This leads to a packet with a Lorentzian energy profile, which in turn has a real space profile of the form
\begin{equation}
\phi (x) = \sqrt{\frac{2 \Gamma}{v_F}} e^{i \epsilon_0 x/v_F} e^{\Gamma x/v_F} \theta(-x) ,
\label{eq:expwp}
\end{equation}
characterized by the set of parameters $\{ \epsilon_0, \Gamma \}$ corresponding respectively to the energy of emission and width in energy of the packet. At low temperature, this exponential wavepacket leads to 
\begin{equation} 
\frac{S_{RL}^{\rm out}(\delta t)}{2 \mathcal{S}_{HBT}} = 1 - e^{-2 \Gamma |\delta t|} .
\end{equation}
Interestingly, this means that the shape of the dip contains relevant information on the properties of the incoming electron packet. 

\begin{figure}
\begin{center}
\includegraphics[height=4.8cm]{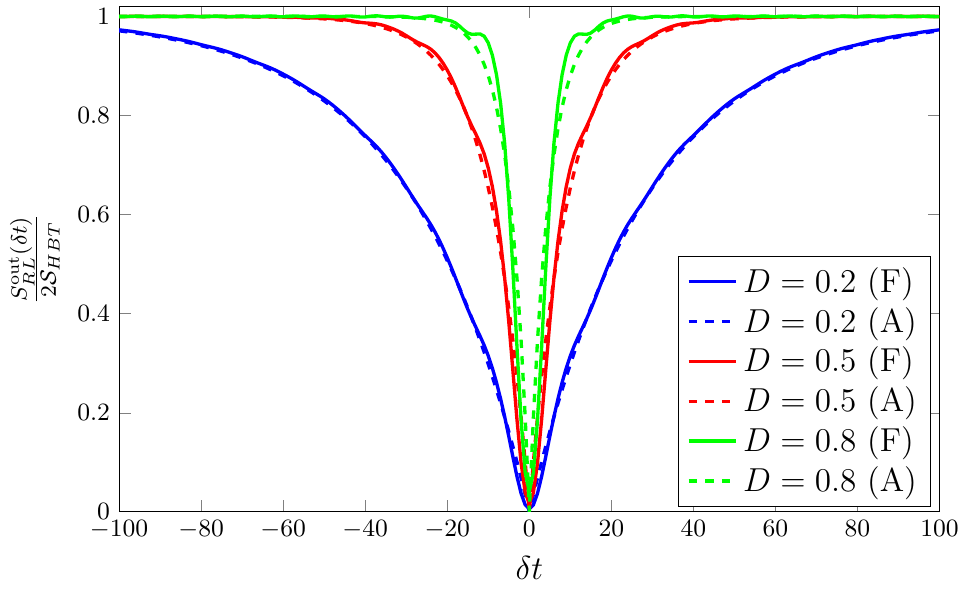}
\end{center}
	\caption{Comparison of the HOM dips obtained as a function of the time difference $\delta t$ from either our analytical calculation (dashed line) or Floquet scattering theory (full line) for different values of the emitter transparency in the case of identical sources.
	 }
\label{fig:symcoll}
\end{figure}

These results can be compared with a Floquet calculation which includes the actual emission process from the SES. A complete detailed description of the source and the corresponding Floquet scattering theory is available in Ref.~\cite{moskalets08,parmentier11}.

Fig.~\ref{fig:symcoll} compares the results of the Floquet calculation for the HOM dip with the analytical formula of Eq.~(\ref{eq:symHOM}). There, the period of the voltage applied to the emitters is $T_0 = 400 \Delta^{-1}$, and the temperature is $\Theta = 0.01 \Delta$, where $\Delta$ is the energy at which electron are emitted into the system. We considered three different values of the transparency $D$ for the SES, corresponding to an electron emission time $\tau = (2\pi/\Delta) (1/D-1/2)$~\cite{parmentier11} itself connected to the width in energy $\Gamma = 1/(2 \tau)$. We observe an HOM dip with different width but maximum contrast (i.e. reaching the minimum zero value at $\delta t =0$) for all three transparencies. Our results show excellent agreement, without any fitting parameters, especially in the low transparency regime where true single electron emission is achieved~\cite{mahe10, grenier11}. The small oscillations present in the Floquet results are typically associated with the ramping up time of the applied square voltage, and are therefore not present in our model of injection.
 
\subsection{Asymmetry and electron-hole collisions} \label{sec:eh}

The previous results can be easily extended to electron-electron collisions of different wavepackets. 
In particular, in the low temperature limit, the expression for the noise dramatically simplifies, and can be written in terms of the overlap of the two incoming wavepackets
\begin{equation}
\frac{S_{RL}^{\rm out}(\delta t)}{2 \mathcal{S}_{HBT}} =
  1- \left| \int dx \; \phi_R(x) \, \phi_L^*(x+ v_F \delta t)\right|^2 ,
\end{equation}
where we assumed for simplicity that the energy content of the wavepacket is above the Fermi level, i.e. $\phi (k) \theta (k-k_F) = \phi (k)$. This last expression is very similar to the one obtained in quantum optics, where the self-convolution of the photon wavepackets sets the shape of the HOM dip \cite{hong87}.

\begin{figure}
\begin{center}
\includegraphics[height=4.8cm]{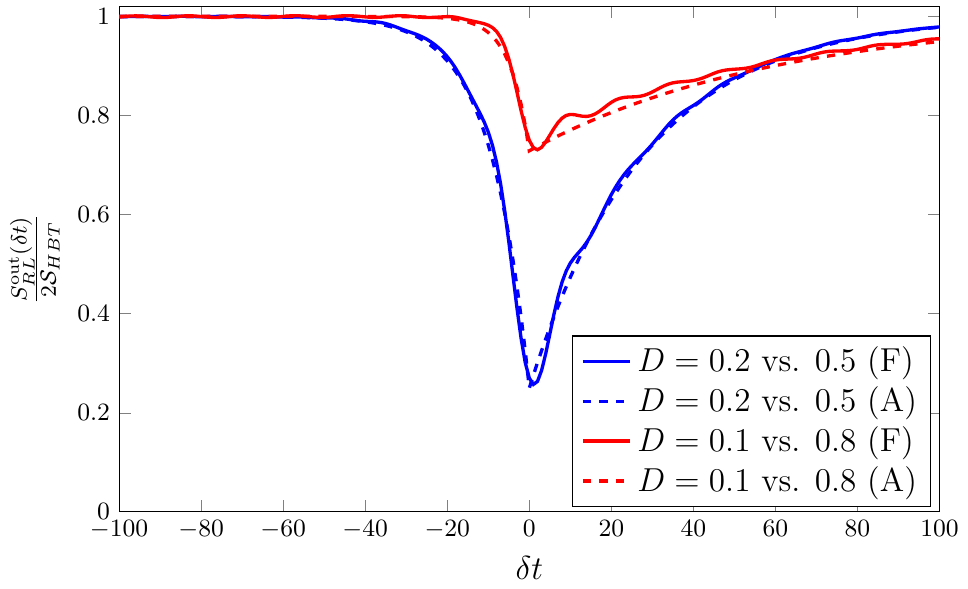}
\end{center}
	\caption{Comparison of the HOM dips obtained as a function of the time difference $\delta t$ from either our analytical calculation (dashed line) or Floquet scattering theory (full line)  in the asymmetric case of sources with different transparencies.
	 }
\label{fig:asymcoll}
\end{figure}

For two exponential wavepackets with different characteristic scales $(\epsilon_\mu, \Gamma_\mu)$, one has at low temperature
\begin{equation}
\frac{S_{RL}^{\rm out}(\delta t)}{2 \mathcal{S}_{HBT}} = 1 -
 \frac{4 \Gamma_R \Gamma_L}{(\Gamma_R + \Gamma_L)^2 + (\epsilon_R - \epsilon_L)^2}  
  \Big[ \theta(\delta t) e^{-2 \Gamma_R \delta t} + \theta(-\delta t) e^{2 \Gamma_L \delta t} \Big] ,
\end{equation}
Like in the symmetric case considered before, this HOM dip has an exponential profile, only with different time constants depending on the sign of $\delta t$. This leads to an asymmetric dip which is moreover characterized by a non-optimal contrast $4 \Gamma_R \Gamma_L/\left[(\Gamma_R + \Gamma_L)^2 + (\epsilon_R - \epsilon_L)^2\right]$, smaller than 1. Such an asymmetry is only possible if the wavepackets have no mirror symmetry in real space. Again, comparison with Floquet scattering theory leads to a very good agreement, as can be seen from Fig.~\ref{fig:asymcoll}, both for the overall asymmetric shape and the value of the contrast.

Hong-Ou-Mandel interferometry in condensed matter devices also offers the intriguing possibility of studying electron-hole collisions, which has no counterpart in regular quantum optics. Injecting a single electron on one incoming edge, and a single hole on the other one, our calculations lead to the following expression for the noise
\begin{eqnarray}
S_{RL}^{\rm out}(\delta t) = -e^2 \mathcal{R} \mathcal{T} \Bigg[
&
\left(
\frac{\int_0^\infty dk
  |\phi_e(k)|^2  (1-f_k)^2}
  {\int_0^\infty dk  |\phi_e(k)|^2 (1-f_k)} \right)^2 +
\left(
\frac{\int_0^\infty dk
  |\phi_h(k)|^2  f_{k}^2}
  {\int_0^\infty dk |\phi_h(k)|^2 f_{k}} \right)^2 \nonumber \\
&
+
 2 \frac{ \left|\int_0^\infty \! dk \,
  \phi_{e}(k) \phi_{h}^*(k) e^{-i k \delta t} f_k (1-f_k)\right|^2}
  {\int_0^\infty \! dk  \, |\phi_e(k)|^2 (1-f_k) \int_0^\infty dk' |\phi_h(k')|^2 f_{k'}}
\Bigg]   ,
\end{eqnarray}
where the first two terms correspond respectively to the HBT contribution of the single electron and the single hole, while the last term is related to interferences between the injected electron and hole.

Several comments are in order at this stage. First, unlike electron-electron collisions, electron-hole interferences contribute positively to the noise, thus leading to an HOM peak rather than a dip. Then, this peak height is conditioned not only upon the overlap of the electron and hole wavepackets through the term $\phi_{e}(k) \phi_{h}^*(k)$, but also upon the product $f_k (1 - f_k)$. This means that the HOM peak vanishes as $\Theta \to 0$, but also that it requires a substantial overlap between electron and hole wavepackets close to the Fermi level in an energy window set by temperature. In particular, the observation of such an HOM peak should require a specific tuning of the source if operated in its optimal regime, or would call for the SES to be driven adiabatically~\cite{jonckheere12}.

\subsection{Experimental results}

The experimental realization of an electronic HOM interferometer in the IQHE occurred recently \cite{bocquillon13} albeit performed in the slightly different regime of filling factor $\nu > 1$, due to technical constraints related to the quantum point contact. The results have several common features with the ones exposed here. Actually, one clearly sees the occurrence of an HOM dip associated with electron-electron collision, and the presence of a flat background contribution (the so-called HBT contribution) that persists for large time difference between emitted electrons. The puzzle with these results is that although one clearly observes an HOM dip, it does not vanish at $\delta t=0$ as predicted for $\nu=1$, therefore signaling interesting effects happening beyond this simple non-interacting picture. 

Indeed, another important difference between photons and electrons is the presence of interactions, and electron quantum optics offers a fascinating playground to explore the emergence of many-body physics. Recent works suggested that interactions may dramatically impact the nature of excitations in integer Hall systems~\cite{levkivskyi08, neder12, berg09, levkivskyi12, kovrizhin11, milletari12, altimiras09, lesueur10, lunde10, degiovanni10}.
This encouraged us to study the case of higher filling factor in the integer quantum Hall regime, and investigate the effect of interactions in the HOM interferometry.


\section{HOM interferometer at filling factor $\nu=2$} \label{sec:nu2}

\begin{figure}
\begin{center}
\includegraphics[height=4.0cm]{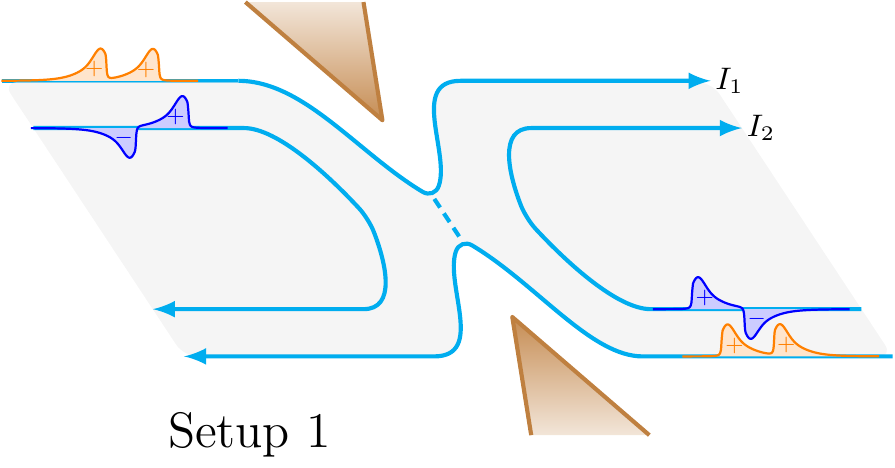}
\includegraphics[height=4.0cm]{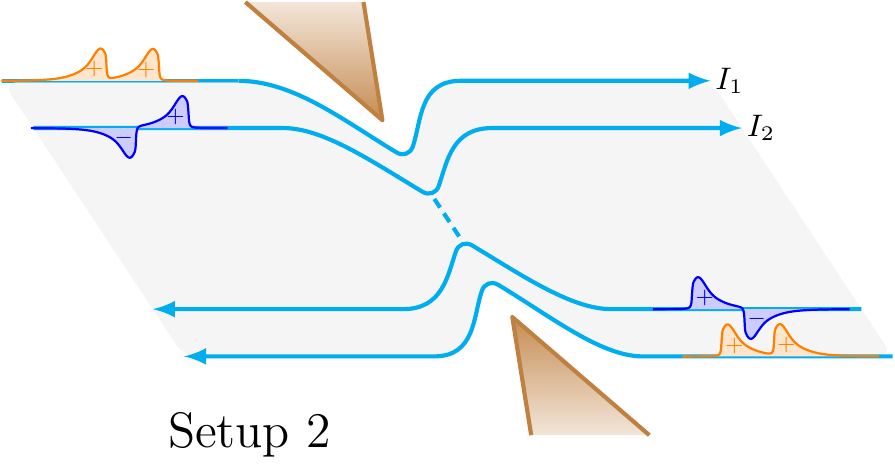}
\end{center}
	\caption{The two possible setups at $\nu=2$: two opposite edge states, each made out of two interacting co-propagating channels, meet at a QPC, and an electronic wavepacket is injected on both incoming outer channels. For setup 1 (left), backscattering occurs for outer channels, as in the experimental device, whereas for setup 2 (right) only inner channels can scatter at the QPC. The fractionalized injected wavepacket is schematically represented through its electron density, revealing the presence of two modes which are each made out of two $\oplus/\ominus$-excitations.
	 }
\label{fig:setupnu2}
\end{figure}

In order to provide a theoretical framework for the experiment, and to further our understanding of the effects of interaction in electronic interferometric setups, we consider now a quantum Hall bar at $\nu = 2$, in the strong coupling regime and at finite temperature. There, each edge state is made out of two co-propagating channels coupled via Coulomb interaction. This is expected to lead to energy exchange between channels, and to charge fractionalization. The two possible setups, referred to as setup 1 and setup 2, correspond respectively to the partitioning of the inner or the outer channel, as shown in Fig.~\ref{fig:setupnu2}.

\subsection{Model and derivation}

Our quantity of interest is the current correlations~\cite{martin92,martin04} measured on the partitioned channel at zero-frequency
\begin{equation}
S^{\rm out} = \int dt dt' \left[ \langle I_s^{\rm out} (t) I_s^{\rm out} (t') \rangle - \langle I_s^{\rm out} (t) \rangle \langle I_s^{\rm out} (t') \rangle \right] ,
\label{eq:defnoise}
\end{equation}
where $s=1,2$ corresponds to the setup considered. Here we used the linear dispersion of the edges, which allows us to compute the noise and thus the outgoing currents at the immediate output of the QPC, without loss of generality. 

Our noise calculations rely on an accurate model of the injection of electrons, their propagation along the edges, and their scattering at the QPC.

\subsubsection{Injection}

The SES is modeled by considering the injection of a single electron in the form of a wavepacket with a definite envelope. Following the lines of our $\nu=1$ treatment, the injection is dealt with by introducing a prepared state similar to the one defined in Eq.~(\ref{eq:injwp}), and compute all average values over this particular state. This consists in a single exponential wavepacket deposited on the outer channel, at a given distance from the QPC, thus mimicking the experimental single electron source in its optimal regime of operation. It is characterized by its injection energy $\epsilon_0$ and energy width $\Gamma$, and is given by Eq.~(\ref{eq:expwp}). Note that the injection always occurs on the outer channel, as this is the most experimentally relevant situation.

\subsubsection{Propagation}

Each edge (labeled $r=R$ and $L$) is made out of two co-propagating channels. These are coupled via Coulomb interaction, which we model as a short-range interaction. 
The outer and inner channels are identified by labels $j=1$ and 2 respectively. Electrons traveling along the edges are more conveniently described in terms of collective bosonic degrees of freedom, following the bosonization formalism~\cite{vondelft98}. It follows that the fermionic annihilation operator $\psi_{j,r} (x,t)$ is written as
\begin{equation}
\psi_{j,r} (x,t) = U_r/\sqrt{2 \pi a}\: e^{i \varphi_{j,r} (x,t)} ,
\label{eq:psiphi}
\end{equation}
where $U_r$ is a Klein factor and $a$ a short distance cutoff parameter, while $\varphi_{j,r} (x,t)$ is the chiral Luttinger bosonic field.

This allows to describe each edge as a chiral Luttinger liquid, with both intra- and inter-channel interactions. The latter can be viewed as a local capacitive coupling between co-propagating channels.
The Hamiltonian is then given by the form $H_0 = H_{\rm kin} + H_{\rm int}$, where
\begin{eqnarray}
	H_{\rm kin} &= \sum_{j=1,2} v_j \frac{\hbar}{\pi} \sum_{r=R,L} \int d x (\partial_x \varphi_{j,r} )^2\\
	H_{\rm int} &= 2 u \frac{\hbar}{\pi} \sum_{r=R,L} \int d x (\partial_x \varphi_{1,r}) (\partial_x \varphi_{2,r}) 
\end{eqnarray}
Here $u$ corresponds to the inter-channel interaction strength, while the intra-channel interaction $U$ has been included in the redefinition of the propagation velocity along the edge, $v_j = v_j^{(0)} + U$.

Upon diagonalization, the fully interacting problem can be recast into a much simpler form using a rotation of angle $\theta$ defined as $\tan (2 \theta) = 2 u /(v_1-v_2)$. In what follows, we focus on the so-called strong coupling regime, corresponding to $\theta=\pi/4$ (and thus to $v_1 = v_2 = v$), as it seems to be the most relevant case from the experimental standpoint \cite{bocquillon13b}. The rotated fields are then given by $\varphi_{\pm,r} =( \varphi_{2,r} \pm \varphi_{1,r})/\sqrt{2}$ and the eigenvelocities reduce to $v_{\pm} = v \pm u$ so that the full Hamiltonian reads
\begin{equation}
H_0 = \frac{\hbar}{\pi} \sum_{r=R,L} \int d x\left[ v_+ (\partial_x \varphi_{+,r} )^2 + v_- (\partial_x \varphi_{-,r} )^2 \right] .
\end{equation}
This Hamiltonian naturally describes two freely propagating collective modes: a fast charged mode and a slow neutral one, traveling along the edge with velocity $v_+$ and $v_-$ respectively. These modes can each be viewed as two separate excitations propagating on the inner and outer channels, and characterized by the charge they carry ($\oplus$ or $\ominus$). 

\subsubsection{Scattering}

The scattering at the QPC is described using a microscopic tunneling Hamiltonian, which for setup 1 takes the form
\begin{equation}
H_{\rm tun} = \Gamma_0 \left[ \psi_{1,R}^\dagger (0) \psi_{1,L} (0) + \psi_{1,L}^\dagger (0) \psi_{1,R} (0)\right] .
\end{equation}

The full Hamiltonian $H=H_0+H_{\rm tun}$ which includes the kinetic, interaction and tunneling parts can be diagonalized~\cite{rufino13} by introducing a new set of fermions $\Psi_{p \pm}$ ($p=A,S$) derived from a refermionization of the bosonic theory as
\begin{equation}
\Psi_{p \pm} (x) = \frac{U_{p \pm}}{\sqrt{2 \pi a}} e^{i \varphi_{p \pm} (x)} ,
\end{equation}
where $\varphi_{p \pm}$ are linear combinations of the bosonic fields introduced in Eq.~(\ref{eq:psiphi})
\begin{eqnarray}
\varphi_{A \pm} &= \pm \frac{(\varphi_{1,R}  - \varphi_{1,L}) \pm ( \varphi_{2,R} - \varphi_{2,L})}{2}\\
\varphi_{S \pm} &= \pm \frac{(\varphi_{1,R} + \varphi_{1,L}) \pm (\varphi_{2,R} + \varphi_{2,L})}{2}
\end{eqnarray}

When expressed in terms of this new set of fermions, the full Hamiltonian $H$ appears quadratic, thus describing a system of non-interacting fermions. This allows us to treat the tunneling at the QPC using a scattering matrix which couples the $A+$ and $A-$ channels:
\begin{equation}
\left( 
\begin{array}{c}
\Psi_{A+}^{\rm out} (t) \\ 
\Psi_{A-}^{\rm out} (t) 
\end{array}
\right)
 = 
\left( 
\begin{array}{cc}
t_0 & -i r_0 \\ 
-i r_0 & t_0
\end{array}
\right)
\left( 
\begin{array}{c}
\Psi_{A+}^{\rm in} (t) \\ 
\Psi_{A-}^{\rm in} (t) 
\end{array}
\right) ,
\label{eq:scatApm}
\end{equation}
where the transmission and reflexion amplitudes $t_0$ and $r_0$ are obtained from the microscopic parameters as $t_0=\sin \varphi$ and $r_0=\cos \varphi$, with $\varphi= - \Gamma_0/(\hbar \sqrt{v_+ v_-})$.

Starting from the expression in terms of the outgoing fermionic degrees of freedom, and using bosonization, refermionization and the scattering matrix of Eq.~(\ref{eq:scatApm}), the current at the output of the QPC can be rewritten as
\begin{eqnarray}
I_{1,R}^{\rm out} (0,t) &= - e \left[ v_F : \left(\psi_{1,R}^{\rm out}\right)^\dagger \psi_{1,R}^{\rm out}   : + u : \left(\psi_{2,R}^{\rm out}\right)^\dagger   \psi_{2,R}^{\rm out}  : \right] (0,t) \nonumber \\
 &= - \frac{e}{2} \left\{ 
   v_+ : \left(\Psi_{S+}^{\rm in}\right)^\dagger  \Psi_{S+}^{\rm in}  :
-  v_- : \left(\Psi_{S-}^{\rm in}\right)^\dagger   \Psi_{S-}^{\rm in}   :
 \right. \nonumber \\
 &   
+ (t_0^2 v_+ - r_0^2 v_-) : \left(\Psi_{A+}^{\rm in}\right)^\dagger  \Psi_{A+}^{\rm in}   : 
+ (r_0^2 v_+ - t_0^2 v_-) : \left(\Psi_{A-}^{\rm in}\right)^\dagger  \Psi_{A-}^{\rm in}   : \nonumber \\
 & \left.
-  i r_0 t_0 (v_+ + v_-) \left[ : \left(\Psi_{A+}^{\rm in}\right)^\dagger  \Psi_{A-}^{\rm in}  : -  : \left(\Psi_{A-}^{\rm in}\right)^\dagger   \Psi_{A+}^{\rm in}   : \right]
\right\} (0,t) ,
\end{eqnarray}
which is in turn recast in terms of the incoming fermionic degrees of freedom as
\begin{eqnarray}
I_{1,R}^{\rm out} (0,t) &= - e \left\{
   \mathcal{T} \left[ v_F  : \left(\psi_{1,R}^{\rm in}\right)^\dagger   \psi_{1,R}^{\rm in}   : 
+ u : \left(\psi_{2,R}^{\rm in}\right)^\dagger   \psi_{2,R}^{\rm in}
\right] 
\right. \nonumber \\
&
+ \mathcal{R} \left[ v_F  : \left(\psi_{1,L}^{\rm in}\right)^\dagger   \psi_{1,L}^{\rm in}   : 
+ u : \left(\psi_{2,R}^{\rm in}\right)^\dagger   \psi_{2,R}^{\rm in}   :  \right] \nonumber \\
&  \left.  
   + i \sqrt{\mathcal{RT}} v_F  : \left[ \left(\psi_{1,R}^{\rm in}\right)^\dagger   \psi_{1,L}^{\rm in}  : 
   -   : \left(\psi_{1,L}^{\rm in}\right)^\dagger  \psi_{1,R}^{\rm in}   : 
\right] \right\} (0,t) ,
 \label{currentrufino}
\end{eqnarray}
where we defined the reflexion and transmission probabilities $\mathcal{R} = r_0^2$ and $\mathcal{T} = t_0^2$.

After some algebra, this allows us to rewrite the current correlations of Eq.~(\ref{eq:defnoise}) in terms of the incoming fermionic degrees of freedom as
\begin{eqnarray}
S^{\rm out} = - e^2 v^2 {\cal R T} \int d t d t' & \left[ \langle \psi_{s,R}^\dagger (t) \psi_{s,R} (t') \rangle \langle \psi_{s,L} (t) \psi_{s,L}^\dagger (t') \rangle \right. \nonumber\\
	&\left. +\langle \psi_{s,L}^\dagger (t) \psi_{s,L} (t') \rangle \langle
	\psi_{s,R} (t) \psi_{s,R}^\dagger (t') \rangle \right] ,
\label{eq:finalnoise}
\end{eqnarray}
where we generalized the approach to both setups $s=1,2$.

Interestingly, this same result can be obtained using a simpler scattering matrix approach, similar to the one used in the previous section for the non-interacting $\nu=1$ case (see Sec.~\ref{sec:nu1formalism}). While such an approach is technically not applicable for interacting fermion fields, it is valid in the present case because both the interaction and the tunneling are purely local~\cite{wahl14}.

\begin{figure}
\begin{center}
\includegraphics[height=4.8cm]{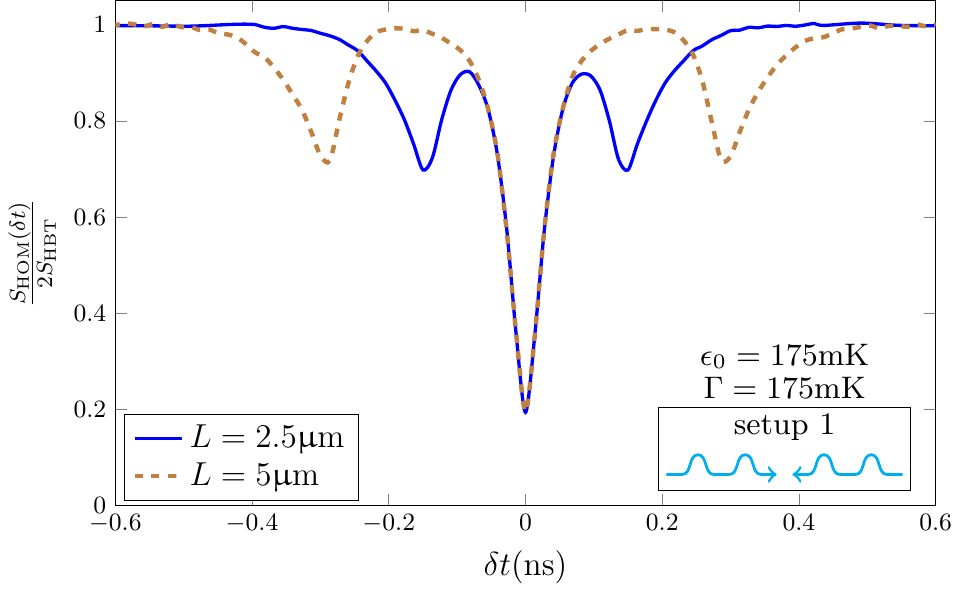}
\includegraphics[height=4.8cm]{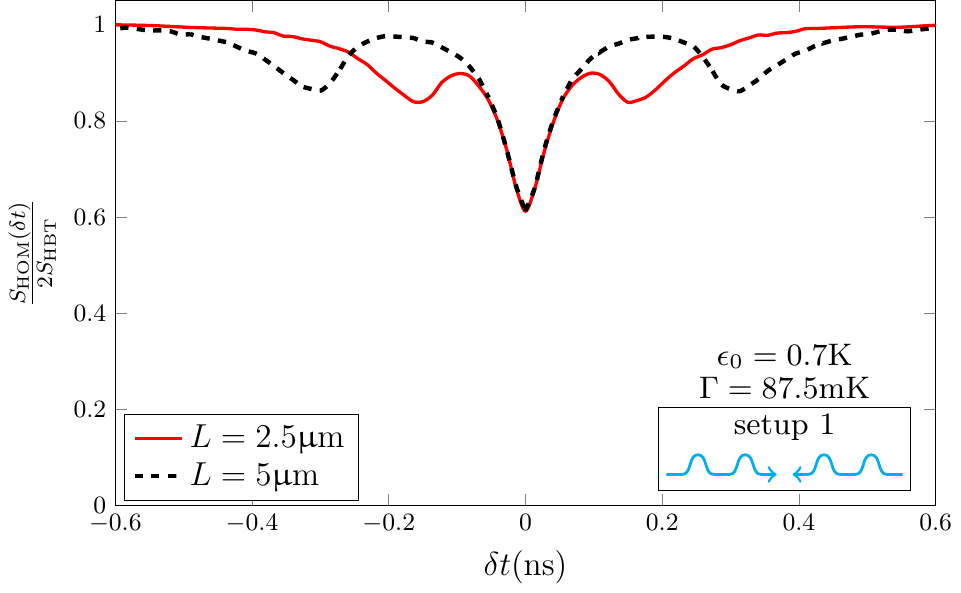}
\end{center}
\caption{Modulus of $S_{\rm HOM}$ in units of $e^2 {\cal R T}$
	as a function of the time delay $\delta t$,
	for setup 1, and two different wavepackets.
	(left) Packets wide in energy with injection energy $\epsilon_0=175 \milli\kelvin$ and energy width $\Gamma = 175 \milli\kelvin$.
	 (right) Energy-resolved packets with injection energy $\epsilon_0=0.7 \kelvin$ and energy width $\Gamma = 87.5 \milli\kelvin$. In both plots, $u=0.5 v$, $\Theta=0.1 \kelvin$, and we considered two different value of the propagation distance $L$.
}
\label{fig:resultssetup1}
\end{figure}

\subsection{Main results}

We now have all the ingredients to compute the noise associated with the HOM configuration, i.e. when injecting a single electron on each incoming outer channel. For sake of simplicity, we proceed with injections at symmetric positions $\pm L$ with respect to the QPC, and consider identical wavepackets on the right and left edge with a time difference $\delta t$. 
Working out explicitly the averages over the prepared state, the expression (\ref{eq:finalnoise}) for the noise becomes
\begin{eqnarray}
S_{\rm HOM} (\delta t) &= - \frac{2e ^2 v^2 {\cal R T}}{(2 \pi a)^4 {\cal N}^2} \mathrm{Re}  
\left\{ \int d y_L d z_L  \phi(L-y_L) \phi^*(L-z_L)  g (0,z_L-y_L) \right. \nonumber \\ 
& \int d y_R d z_R \phi(L+y_R) \phi^*(L+z_R)  g(0,y_R-z_R)  \times \int d \tau \mathrm{Re} \left[ g(\tau,0)^2 \right] \nonumber\\
& \left.   \int  dt \left[ \frac{h_s(t;y_L,z_L)}{h_s (t+\tau;y_L,z_L)} \frac{h_s (t+\tau-\delta t;-y_R,-z_R)}{h_s (t-\delta t;-y_R,-z_R)}   - 1 \right] \right\} ,
		 \label{eq:HOMnoise}
\end{eqnarray}
with the wavepacket envelope $\phi (x) = \sqrt{\frac{2 \Gamma}{v}} e^{i \epsilon_0 x/v} e^{\Gamma x/v} \theta(-x)$, and normalization ${\cal N}= \langle \phi | \phi \rangle$. The functions $g$ and $h_s$ are obtained from the Green's function of the bosonic $\varphi_{\pm r}$ fields, and are defined as 
\begin{eqnarray}
g(t,x)&= \left[ \frac{\sinh \left( i \frac{\pi a}{\beta v_+} \right)}{\sinh \left(  \frac{ia + v_+ t - x}{\beta v_+ /\pi} \right)}  \frac{\sinh \left( i \frac{\pi a}{\beta v_-} \right)}{\sinh \left( \frac{ia + v_- t - x}{\beta v_- /\pi} \right)} \right]^{1/2} , \\
h_s (t;x,y) &= \left[ \frac{\sinh \left( \frac{ia - v_+ t + x}{\beta v_+ / \pi} \right)}{\sinh \left(\frac{ia + v_+ t - y}{\beta v_+/\pi} \right)} \right]^{\frac{1}{2}} \left[ \frac{\sinh \left( \frac{ia - v_- t + x}{\beta v_- / \pi} \right)}{\sinh \left( \frac{ia + v_- t - y}{\beta v_- /\pi} \right)} \right]^{s-\frac{3}{2}} .
\end{eqnarray}
where $s=1,2$ is the setup considered. The noise is obtained numerically via multidimensional integration handled with a quasi Monte Carlo algorithm using importance sampling~\cite{hahn05}.

\begin{figure}
\begin{center}
\includegraphics[height=4.8cm]{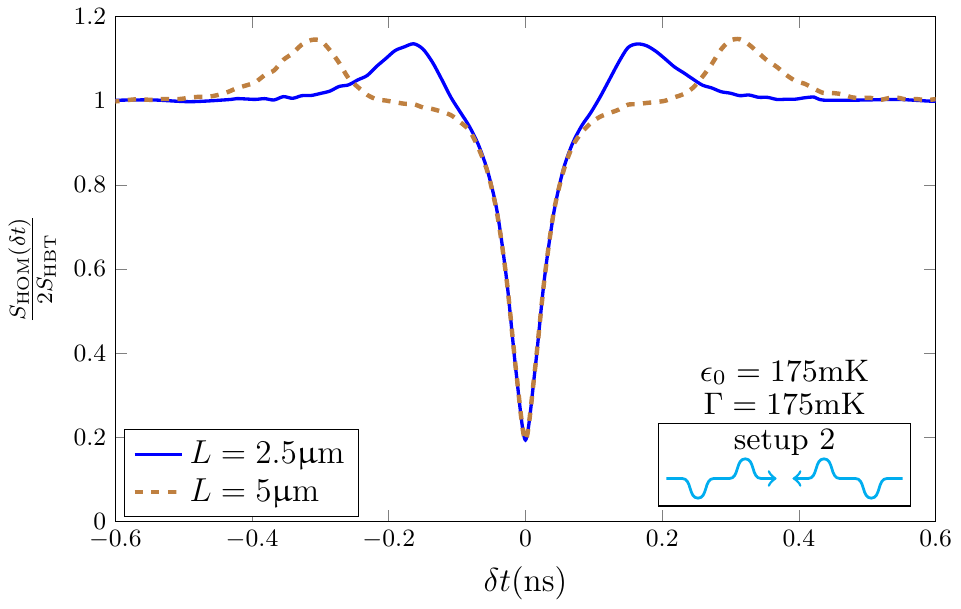}
\includegraphics[height=4.8cm]{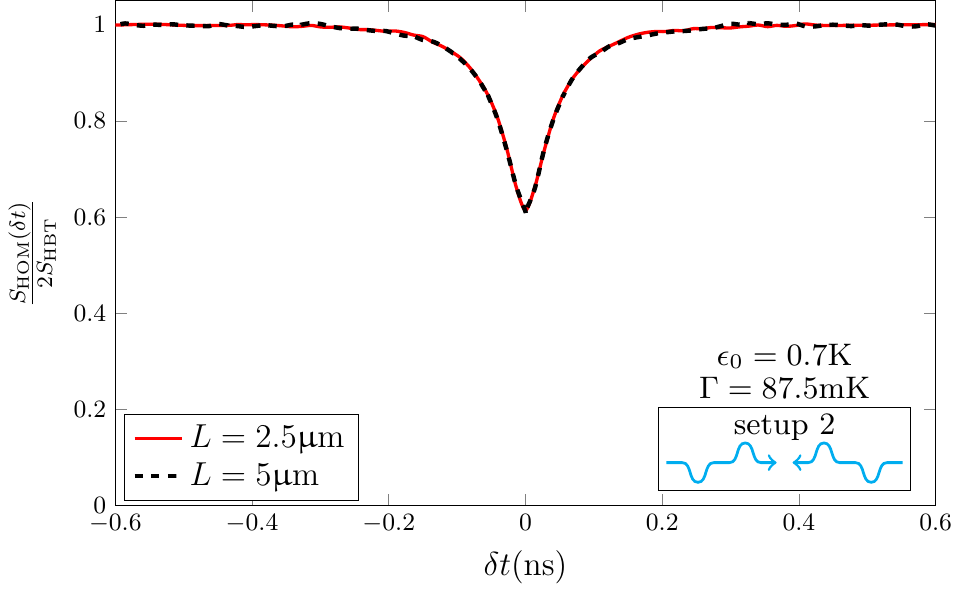}
\end{center}
	\caption{Modulus of $S_{\rm HOM}$ in units of $e^2 {\cal R T}$
	as a function of the time delay $\delta t$,
	for setup 2, and two different wavepackets.
	(left) Packets wide in energy with injection energy $\epsilon_0=175 \milli\kelvin$ and energy width $\Gamma = 175 \milli\kelvin$.
	 (right) Energy-resolved packets with injection energy $\epsilon_0=0.7 \kelvin$ and energy width $\Gamma = 87.5 \milli\kelvin$.  In both plots, $u=0.5 v$, $\Theta=0.1 \kelvin$, and we considered two different value of the propagation distance $L$.
}
\label{fig:resultssetup2}
\end{figure}

Our computations of the output current correlations as a function of the time delay between right- and left-moving injected electrons reveal three characteristic signatures. Away from these three features, $S_{\rm HOM}$ saturates at twice the HBT noise $S_{\rm HBT}$ as the  electrons injected on the two incoming arms scatter independently at the QPC without interfering. The interference patterns are provided in Figs.~\ref{fig:resultssetup1} (for setup 1) and~\ref{fig:resultssetup2} (for setup 2) for a given set of parameters, and the various structures can be interpreted in terms of the different excitations propagating along the partitioned edge channel. Indeed, after being injected, the electron fractionalizes into a fast and a slow mode. The fast mode is charged and made out of two $\oplus$ excitations. The slow mode, on the other hand, is neutral and composed of a $\oplus$ excitation propagating along the outer channel and a $\ominus$ excitation traveling along the inner one. 

The most striking signature appears at a time delay $\delta t=0$ in the form of a central dip. This dip probes the interference of both fast and slow right-moving excitations with their left-moving counterparts, i.e. of colliding excitations which have the same velocity and charge. These interfere destructively, resulting in a reduction of the noise (in absolute value) and thus a dip. While its depth strongly correlates with the energy resolution of the injected wavepackets, the dip depends very little on the setup considered, which suggests that the interference mechanism is the same for $\oplus/\oplus$ and $\ominus/\ominus$ collisions.

\begin{figure}
\begin{center}
\includegraphics[height=4.8cm]{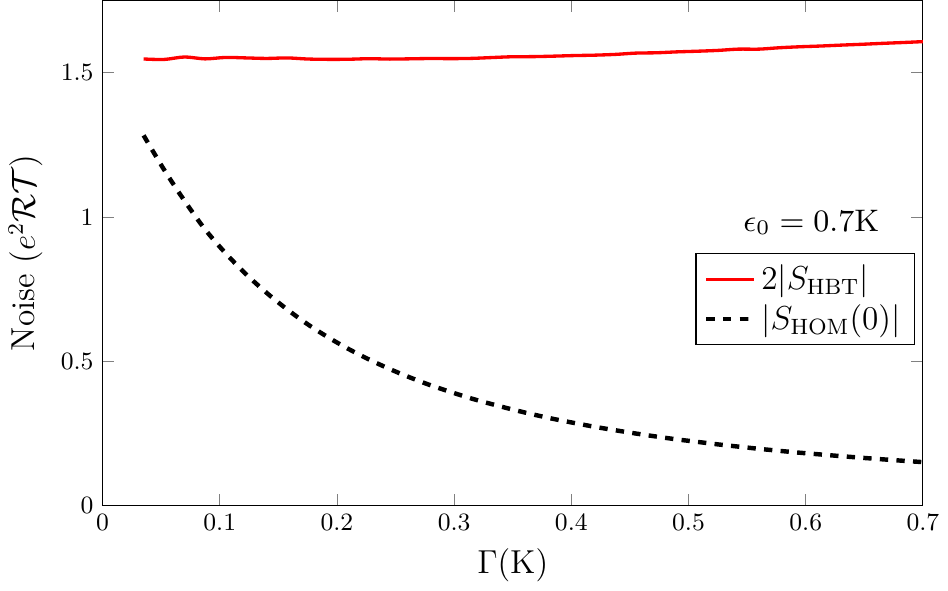}
\end{center}
\caption{Comparison of the HBT contribution with the HOM noise at zero time delay as a function of the energy width of the incoming wavepackets. Here the injection energy is $\epsilon_0 = 0.7 \kelvin$, the interaction parameter $u = 0.5 v$ and the temperature $\Theta = 0.1 \kelvin$. The HBT contribution is almost constant as a result of the competition between the creation of particle-hole pairs (favored as the resolution increases~\cite{degiovanni09}) and their anti-bunching with thermal excitations at the output of the QPC~\cite{bocquillon12}.
}
\label{fig:HOMcontrast}
\end{figure}

As observed in the experiment, the central dip never quite reaches zero in our calculations, in striking contrast with the $\nu=1$ case~\cite{bocquillon13}. This is actually a probing tool of the degree of indistinguishability of the excitations colliding at the QPC.  Because of the strong inter-channel interaction, some coherence of the injected object is lost in the co-propagating channels which do not scatter, and this Coulomb-induced decoherence is responsible for the dramatic reduction of contrast of the HOM dip. For a fixed injection energy, this effect becomes more pronounced as the energy width of the wavepacket is reduced (or alternatively as the emission time increases), as depicted in Fig.~\ref{fig:HOMcontrast}. Indeed, the more resolved in energy a wavepacket is, the more it is subject to decoherence~\cite{ferraro14}, leading to a net reduction of the contrast.

Smaller satellite structures also appear in the noise, but at finite delay $\delta t$. These emerge symmetrically with respect to the central dip at positions $\delta t= \pm 2 L u /(v^2-u^2)$. The shape and depth of these features depend on the energy resolution of the wavepacket and vary critically between setups, manifesting as dips for setup 1, but peaks for setup 2. They show a non-trivial dependence on the wavepacket energy content, being more pronounced for packets wide in energy but vanishingly small for well-resolved ones.

These structures appear as a consequence of interference between excitations that have different velocities, when a fast and a slow-moving excitations reach the QPC at the same time. For setup 1, this corresponds to two colliding $\oplus$ excitations, which interfere destructively resulting in dips. 
For setup 2, however, the satellite peaks are associated with the collision of oppositely charged excitations, which leads to constructive interference, and thus to a peak. This is reminiscent of the electron-hole interferometry considered in the $\nu=1$ case in Sec.~\ref{sec:eh}.

The lateral dips are asymmetric with a depth less than half the one of the central dip. These properties of asymmetry and reduced contrast are reminiscent of the behavior encountered in the non-interacting $\nu=1$ case when colliding packets of different shapes (see Fig.~\ref{fig:asymcoll}). Here, it is a consequence of the velocity mismatch between interfering excitations.

\begin{figure}
\begin{center}
\includegraphics[height=4.8cm]{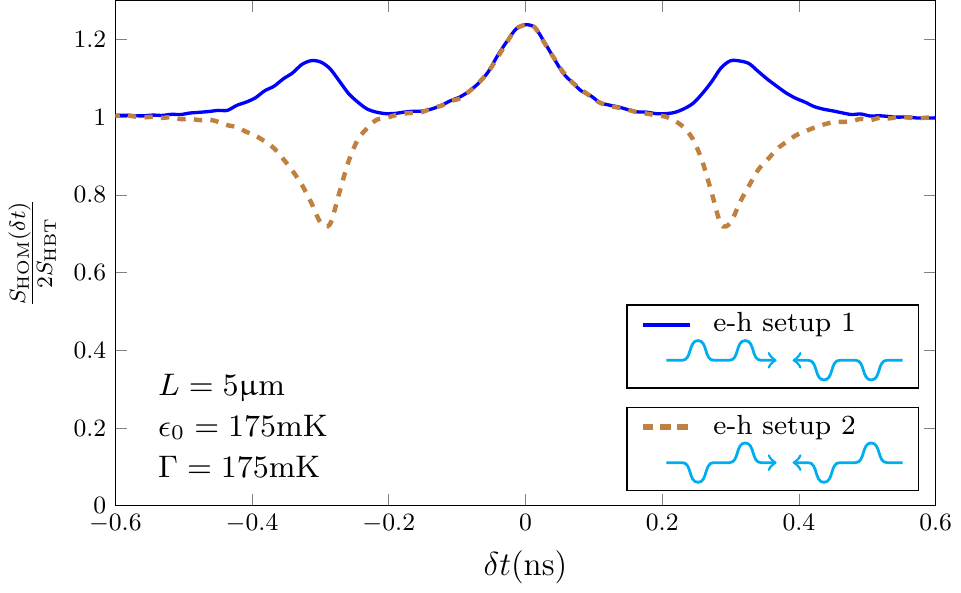}
\end{center}
\caption{Electron-hole HOM interference: an electron has been injected
	on the right moving arm and a hole on the left moving one. Noise obtained for symmetric wavepackets with injection energy $\epsilon_0 = \pm 175 \milli\kelvin$, energy width $\Gamma = 175\milli\kelvin$, interaction parameter $u=0.5 v$ and temperature $\Theta=0.1 \kelvin$.
}
\label{fig:nu2eh}
\end{figure}

Finally, our approach can also be extended to the case of electron-hole collisions. As in the electron-electron interferometry, this leads to three signatures in the noise (see Fig.~\ref{fig:nu2eh}). First, a central peak appears at $\delta t =0$ for both setups, corresponding to the constructive interference of a $\oplus$ with a $\ominus$ excitation. Then, satellite features are also present, manifesting as peaks for setup 1 (produced by interfering oppositely charged excitations) and dips for setup 2 (probing the interference of same charge excitations).

%
%


\section{Beyond the integer case: unsolved problems} \label{sec:upon}

Interactions dramatically change the nature of the excitations, and the HOM interferometry offers the possibility to probe the incoherent mixture of fractionalized electronic excitations induced by Coulomb interactions. A natural extension of this work consists in studying a system where the ground state itself is a strongly correlated state of matter: the fractional quantum Hall effect (FQHE). There, one would be dealing not with electrons, but with single quasiparticles with fractional charge and statistics which should lead to dramatically new physics.

This constitutes a challenge at various levels, as a lot of open and fascinating questions remain.
\begin{itemize}
\item Can we emit controlled single quasiparticles in the system? What would be the nature of the quasiparticle injector?

This is a fundamental prerequisite for the realization of HOM interferometry. The current design of single electron source cannot be readily extended to emit quasiparticles in the FQHE as it could only operate in the strong backscattering regime. Recently, some of us proposed an antidot-based device susceptible to work as an on-demand single quasiparticle source with little to no charge fluctuations \cite{ferraro15}. However a purely Hamiltonian description of such a non-equilibrium system is still lacking.

\item Is a perturbative treatment in tunneling sufficient?

Standard calculations implying a QPC in the fractional regime rely on a perturbative treatment in powers of the tunneling constant (typically up to second order). Whether such calculations would be sufficient to capture the physics involved in fractional HOM interferometry deserves to be explored.

\item Do quasiparticles show bunching? Are there signatures of non-trivial statistics in the HOM noise signal?

The link between the measurement of low frequency noise correlations and the statistics of the carriers is well known. HOM interferometry with photons or electrons allows to probe the statistics through second order coherence, whether this is also enough to access the fractional statistics of quasiparticles is still under debate.


\end{itemize}


\section{Conclusions} \label{sec:conclusion}

To conclude, we studied the HOM interferometer in the integer quantum Hall regime. In the non-interacting $\nu=1$ case, we proved that the zero-frequency current correlations exhibit a dip when two electrons collide or a peak for electron-hole collisions, with a shape tied to the characteristics of the injected wavepackets. Our analytic calculations agreed well with Floquet scattering theory which allows to consider more accurately the experimental single electron source.

In the $\nu=2$ case, we showed that the HOM dip survives but that the strong coupling between co-propagating channels accounts for a sensible loss of contrast, as observed in the experiment. This reduction is a direct consequence of decoherence and strongly depends on the energy content of the colliding electronic wavepackets. 
Moreover, this situation leads to a richer interference pattern, with the presence of asymmetric side dips and peaks related to the interference of fast and slow modes. 

In a natural extension of this work, we discussed the case of fractional HOM interferometry, pointing out the main unsolved problems and open questions that need to be tackled before such a fascinating possibility can be explored.

\ack

We are grateful to D. Ferraro, G. F\`eve, B. Pla\c cais and P. Degiovanni for useful discussions. This work was granted access to the HPC resources of Aix-Marseille Universit\'e financed by the project Equip@Meso (Grant No. ANR-10-EQPX-29-01). It  has been carried out in the framework of project "1shot" (Grant No. ANR-2010-BLANC-0412) and benefited from the support of the Labex ARCHIMEDE (Grant No. ANR-11-LABX-0033) and of the AMIDEX project (Grant No. ANR-11-IDEX-0001-02), all funded by the “investissements d’avenir” French Government program managed by the French National Research Agency (ANR).

\section*{References}

\bibliographystyle{iopart-num}
\bibliography{1shot_biblio}

\end{document}